\PassOptionsToPackage{hyphens}{url}
\PassOptionsToPackage{dvipsnames}{xcolor}

\documentclass[conference]{IEEEtran}

\usepackage{amsmath}
\usepackage{amssymb}
\usepackage{physics}
\usepackage{dsfont}
\usepackage{mathrsfs}
\usepackage{wasysym}

\usepackage[hyphens,obeyspaces,spaces]{url}
\usepackage[utf8]{inputenc}
\usepackage{array, multirow}
\newcolumntype{P}[1]{>{\centering\arraybackslash}p{#1}}
\usepackage{tikz}
\usepackage{listings}
\usepackage{makecell}
\usepackage{subcaption}
\usepackage{flushend}
\usepackage{fancyhdr}
\usepackage{listings}
\usepackage{cprotect}
\usepackage{environ}
\usepackage{authblk}
\usepackage{mathtools}
\usepackage{svg}
\usepackage{xcolor}
\usepackage{MnSymbol,wasysym}
\usepackage{caption}
\usepackage{pifont}

\usepackage{url}
\usepackage{amsmath,amsfonts}
\usepackage{graphicx}
\usepackage{balance}
\usepackage{makecell}
\usepackage[ruled,vlined]{algorithm2e}
\usepackage{todonotes}
\usepackage{graphicx}
\usepackage{tabularx}
\usepackage{subcaption}
\newcommand{\ignore}[1]{}

\newcommand{\brpki}{{\small{{BRP}}}}

\definecolor{ao}{rgb}{0.0, 0.5, 0.0}

\newcommand*\circled[1]{\tikz[baseline=(char.base)]{
            \node[shape=circle,draw,inner sep=0pt,minimum size=12pt] (char) {\small #1};}}

\begin{document}

\title{Byzantine-Secure Relying Party for Resilient RPKI} 

\author{Jens Frieß$^{\dagger\natural}$\qquad Donika Mirdita$^{\dagger\natural}$\qquad Haya Schulmann$^{\ddagger\natural}$\qquad Michael Waidner$^{\dagger\S\natural}$\vspace{2mm}\\
{\small
$^\dagger$ Technische Universität Darmstadt \qquad $^\ddagger$ Goethe-Universität Frankfurt \qquad $^\S$ Fraunhofer SIT \qquad
$^\natural$ ATHENE} \\
}

\maketitle

\begin{abstract}

BGP is a gaping hole in Internet security, as evidenced by numerous hijacks and outages. The significance of BGP for stability and security of the Internet has made it a top priority on the cyber security agenda of nation states, with the US government, in particular CISA, FCC, and other federal agencies leading the efforts\footnote{\url{https://www.cisa.gov/news-events/news/most-important-part-internet-youve-probably-never-heard}}. 

To protect against prefix hijacks, Resource Public Key Infrastructure (RPKI) has been standardized. Yet,  RPKI validation is not widely supported. To enjoy the security guarantees of RPKI validation, the networks need to install a new component, the relying party validator, which fetches and validates RPKI objects and provides them to border routers. However, recent works showed that relying parties experience failures when retrieving RPKI objects and are vulnerable to different attacks, all of which can disable RPKI validation. Therefore even the few adopters are not necessarily secure.

We make the first proposal that significantly improves the resilience and security of RPKI validation. We develop \brpki, a Byzan-tine-secure relying party implementation. In \brpki\ the relying party nodes redundantly validate RPKI objects and arrive at a global consensus through a voting process.  \brpki\ provides an RPKI equivalent of public DNS, removing the need for networks to install, operate, and upgrade their own relying party instances while avoiding the need to trust operators of \brpki\ nodes.

We show through simulations and experimental evaluations that \brpki, as an intermediate RPKI service, results in less load on RPKI publication points and a robust output despite RPKI repository failures, jitter, and attacks. We engineer \brpki\ to be fully backward compatible and readily deployable - it does not require any changes to the border routers and the RPKI repositories. \brpki\ enables independent verification by users of its correct operation.

We demonstrate that \brpki\ can protect many networks transparently, with either a decentralized or a centralized deployment. \brpki\ can beset up as a network of decentralized volunteer deployments, similarly to NTP and TOR, where different operators participate in the peering process with their relying party node, and provide resilient and secure relying party validation to the Internet. \brpki\ can also be hosted by a single operator as a centralized service, e.g., on one cloud or CDN, and even provides RPKI validation benefits when hosted on just a single network.

We make the code of \brpki\ and the evaluation data public\footnote{\url{https://anonymous.4open.science/r/resrpki/}}. 

\end{abstract}

\section{Introduction}
One of the most glaring security vulnerability of today's Internet infrastructure is the insecurity of the Border Gateway Protocol (BGP). Border routers exchange BGP announcements to establish routes between the Autonomous Systems (ASes) that make up the Internet. BGP's susceptibility to configuration errors and attacks is the cause for major Internet outages~\cite{ripencc2008,BGPmon_Indosat,BGPmon_turk} and traffic hijacks~\cite{china:telecom,ballani2007study,fb:out,u:tube,mitm:threat,vervier2015mind}. 

For example, a recent BGP hijack allowed attackers to steal 235.000\$ in cryptocurrency from a crypto transfer service by hijacking the AS number of the AS hosting the service\footnote{\url{https://arstechnica.com/information-technology/2022/09/how-3-hours-of-inaction-from-amazon-cost-cryptocurrency-holders-235000/}}. BGP hijack incidents are common and often lead to large scale outages. Over the past 10 years, prefix hijacks have become more refined and targeted.

{\bf Resource Public Key Infrastructure.} A crucial step in today's agenda for securing BGP is deploying the Resource Public Key Infrastructure (RPKI) \cite{rfc6481}. RPKI binds IP address blocks to ``owner'' ASes via cryptographic signatures, which are then stored in RPKI objects in public RPKI repositories. RPKI enables ASes to discard BGP announcements originating in an adversarial or a misconfigured router, that claims to own an IP address block that belongs to another Internet origin (``IP-prefix-hijacks''). 
To filter bogus BGP announcements, networks need to set up a relying party validator. Relying parties fetch and validate the certificates and other RPKI objects from the public RPKI repositories, creating Validated ROA Payloads (VRPs). Routers retrieve the VRPs from the caches of the validators and use them to identify and filter bogus BGP announcements with Route Origin Validation (ROV). 

{\bf Low RPKI adoption.} RPKI was standardized more than a decade ago, in 2011, but despite its significance, RPKI's deployment is discouragingly slow. Although almost 50\% of all Internet prefixes are covered with ROAs, the validation of BGP announcements against the ROAs is done only by a small number of Internet networks. The numbers depend on the concrete study. According to RoVista measurement \cite{rovista} about 12.3\% enforce ROV, i.e., fetch the ROAs and other RPKI objects and validate the BGP announcements against them. Other studies report up to 30\% support ROV \cite{testart2020filter,cloudflare,apnic}.

{\bf Obstacles and hurdles to RPKI deployment.} There are a number of factors preventing wide deployment of RPKI. 

$\triangleright$ First, the networks have to set up and configure a relying party on a well-connected network to ensure that the RPKI objects can always be retrieved and validated. Deploying a relying party is a complex and error-prone manual task. The Pareto Principle, i.e., the ``80:20 rule'' \cite{sanders1987pareto}, suggests that most of the desired outcomes are achievable with significantly less effort than needed to realize all one desires, with increasing effort resulting in diminishing returns. Defenses should follow the Pareto Principle: the cost and effort of deploying a security mechanism should be minimized and security benefits are maximized, which is currently not the case with RPKI validation. Consequently, the motivation to deploy RPKI is low.

$\triangleright$ Another issue is that the relying parties often fail to connect to the repositories to fetch the RPKI objects \cite{hlavacek2023beyond}. No access to a complete, correct, and up-to-date set of RPKI objects exposes the networks to BGP prefix hijack attacks. The reason is that when a covering ROA for a given prefix is not available, the RPKI validation of BGP announcements for that prefix results in status "unknown". As a result, although those prefixes are signed in RPKI, ROV will not filter hijacks for them. Namely, currently even the fairly few adopters are not necessarily secure. A security mechanism, whose protection can be removed, e.g., by creating a load on the RPKI repositories - a completely realistic and practical attack - creates a false sense of security, leading, in fact, to a less secure Internet.

$\triangleright$ Another significant obstacle is that about 56\% of all the relying party validators on the Internet are vulnerable to some form of Denial of Service, stalling or root poisoning attacks \cite{van2022rpkiller,stalloris,ndss:2024}. Consequently, operators do not rush to introduce another vulnerable component to their networks.

As the adoption of RPKI proceeds, any inconsistency, vulnerability, or misconfiguration in RPKI will have a greater impact on the Internet stability, since increasingly more networks may be affected. These failures and vulnerabilities also reduce the motivation to deploy RPKI.

{\bf Byzantine-secure relying party.} To resolve these issues in RPKI and facilitate wide RPKI adoption we propose \brpki, a new resilient intermediate relying party layer for RPKI. \brpki\ uses a centralized setup with multiple relying party instances and runs a Byzantine agreement protocol between them to reach consensus on RPKI objects. \brpki\ achieves good synchronization even against strong adversaries that can attack, corrupt or control some of the relying parties. \brpki\ is based on the concept of consensus in the presence of adversarial (Byzantine) behaviour from the distributed computing research \cite{dolev1986reaching}. The idea is that a number of relying parties periodically fetch and validate the RPKI objects from the distributed repositories (hosted on RPKI publication points) on the Internet. Then they run a consensus algorithm among them to identify the most up to date and complete set of RPKI objects. The algorithm eliminates faulty or missing or outdated RPKI objects. We develop the \brpki\ system to be efficient, requiring low communication overhead to achieve synchronization. The border routers on the Internet retrieve these validated RPKI objects (VRPs) over the RTR protocol from \brpki\ like they previous did from the local relying party instances. \brpki\ is compatible with the existing routing and RPKI infrastructures and can be readily used by the border routers. There are no requirements on the part of the border routers and no requirements on the part of RPKI repositories for communication with our \brpki. 

{\bf Contributions.} We identify the key element in RPKI's architecture that exposes it to failures and attacks: the relying party functionality for fetching and validating the RPKI objects. During failures or attacks the relying parties cannot provide a complete set of VPRs to the routers, which in turn cannot enforce ROV. We make the first proposal to significantly improve the resilience and security of the relying party validation. To that end we develop \brpki: a ``Byzantine consensus''- based relying party implementation. \brpki\ is carefully engineered to support the following design goals: 

$\triangleright$ \brpki\ ensures validation of RPKI objects and correct and resilient generation of VRPs, which the border routers can use to protect their networks against BGP prefix hijacks. We outline the specific issues in RPKI that make the design with a Byzantine consensus challenging and show how we overcome them in our design. 

$\triangleright$ We engineer \brpki\ to be fully backwards compatible. A major obstacle towards deployment of new mechanisms is the requirement to modify the existing infrastructure, software or protocols. \brpki\ does not require any changes to the existing systems and devices, it is readily deployable for communication with the RPKI repositories and the border routers.

$\triangleright$ We analyze the security of our proposal and show experimentally, on real Internet traffic as well as on empirically derived attack datasets, that \brpki\ is not only more secure but also more resilient and performs better in contrast to existing vanilla relying parties. We demonstrate using current statistics and empirical measurements that \brpki\ minimizes the load on repositories and the network compared to individually deployed relying parties and extrapolate this reduction to full RPKI deployment. We show that \brpki\ would produce only 1-1.5\% of the estimated RPKI traffic.

$\triangleright$ We envision \brpki\ to be a network of decentralized volunteer deployments, similarly to NTP and TOR \cite{alsabah2016performance,moura2024deep}, where different operators participate in the peering process with their relying party node, and provide resilient and secure relying party validation to the Internet. \brpki\ can also be hosted by a single operator as a centralized service, e.g., on one cloud or CDN, and even provides RPKI validation benefits when hosted on just a single network.

$\triangleright$ We make the code of \brpki\ as well as the datasets, that we collected for the evaluations in this work public \footnote{\url{https://anonymous.4open.science/r/resrpki/}}. Our public repository also contains a readme file that explains the setup options (on local network, or on a cloud platform, or as a distributed volunteer network) and provides instructions how to realize them.

{\bf Organization.} We put our research in context of related work in Section \ref{sc:works}. We then review RPKI, explain key problems in Section \ref{sc:rpki}, and discuss RPKI-related challenges inherent in the development of a Byzantine-secure relying party in Section \ref{sc:challenge}. 
We provide an overview of Byzantine agreement in Section \ref{sc:byzantine}. In Section \ref{sc:brp} we introduce the design and implementation of a Byzantine-secure relying party, we call \brpki. We analyze security in Section \ref{sc:security}, and perform evaluations to demonstrate correctness, resilience, and performance in Section \ref{sc:eval}. We conclude this work in Section \ref{sc:conclusions}.

\section{Related Work}\label{sc:works}

There were many tools proposed for detecting prefix hijacks, such as PHAS \cite{lad2006phas}, Argus \cite{xiang2011argus}, ARTEMIS \cite{sermpezis2018artemis}, HEAP \cite{schlamp2016heap}. These proposals use heuristics, and not cryptographic mechanisms for validating BGP announcements, therefore allowing sophisticated adversaries to evade monitoring \cite{milolidakis2021poster}. A completely different approach to securing the inter-domain routing postulates to replace BGP with SCION \cite{zhang2011scion}. Since it does not seem that BGP will be replaced any time soon, the SIDR (Secure Inter-Domain Routing) working group designed RPKI \cite{RFC6480} to enable networks to filter BGP hijacks.

 \cite{gilad2018perfect} identified that one of the hurdles in RPKI is the manual verification of prefix ownership. To address this obstacle \cite{hlavacek2020disco} proposed to automate the certification of prefixes in RPKI, a task which is traditionally done by the Regional Internet Registries (RIR). The idea is that certification should be carried out automatically based on de-facto ownership of prefixes instead of legal ownership.

A lack of deployment of path validation mechanisms motivated extensions to RPKI, to develop an alternative mechanism to a full fledged BGPsec \cite{cohen2015one,cohen2016jumpstarting}. Their proposal, called "path-end validation", attempts to ensure that the last hop on the BGP path is valid. 

Recently \cite{morillo2021rov++} found a limitation in ROV filtering performed by the border routers and showed that ROV does not effectively block subprefix hijacks and non-routed prefix hijacks. \cite{morillo2021rov++} proposed to extend ROV by modifying the border routers to retrieve additional policy modules. The demonstrate that their proposal also increases security benefits of ROV in partial deployment. Our mechanism does not require any changes to BGP routers nor RPKI components, and do not require installing new devices on the networks.

Different to all previous work, we focus on the relying party functionality. We trace key problems in the current retrieval and validation of RPKI objects to load, failures and vulnerabilities in relying party software packages. We make the first proposal to extend the relying party by making it resilient to attacks and failures with a Byzantine agreement, i.e., a ``Byzantine secure''-relying party functionality, which we call \brpki. Our design of \brpki\ supports a global setup on a cloud, a local setup on a specific network, as well as a distributed setup of multiple relying parties on different networks, all running our \brpki\ consensus. We analyze \brpki\ and demonstrate through comprehensive experimental evaluations that \brpki\ provides a practical, resilient and a secure enhancement of the RPKI validation functionality, bootstrapping ROV filtering of the border routers with a correct and consistent set of VRPs.

\section{RPKI and Practical Problems}\label{sc:rpki}
We provide an overview of RPKIs and discuss key problems which we aim to resolve with our proposal Byzantine-secure relying party.

\subsection{A Primer on RPKI}

RPKI is composed of a hierarchically distributed set of repositories called Publication Points (PPs) and validator software called Relying Party (RP). 
RPKI uses certificates issued by Regional Internet Registries (RIRs) as anchors to establish a chain of trust for the validation of all the objects in the repositories. RIRs publish Trust Anchor Locators (TALs) \cite{rfc6490} files that point towards the RIR RPKI root certificate. The certificate then points to the location of the repository data for that RIR. 

{\bf Publication Points (PPs)} serve objects for one or more Certificate Authorities (CAs). CAs are the logical and legal units of RPKI data management. Each CA represents a network resource-owning entity that has consented to include its data in the RPKI ecosystem. CAs can decide to delegate RPKI data management to the resource provider, e.g., the RIR, thus using {\it hosted RPKI} mode, where the resource provider hosts the RPKI data of this CA on its own repository, or it can administer its own unique repository in a {\it delegated RPKI} mode. Each new RPKI CA must provide their parent with a certificate that points to the location of that CA's RPKI objects. The RPKI repositories, that are hosted on the publication points, contain several types of objects to ensure data completeness and integrity. The short-lived manifest files contain an ever-updating list of the hashes of current objects in the repository, certificate revocation lists, and X.509 certificates of child CAs. 
All RPKI objects are cryptographically signed by the owner CA's private keys, and any updates to the CA repository content requires signature regeneration.

{\bf Relying Party (RP)} is a software component that periodically fetches, parses and cryptographically validates RPKI objects. RPs begin every validation round at predefined refresh intervals, with the TALs of the RIRs, which point to the RIR repositories, which in turn contain certificate pointers to child CAs and so on. This process allows the RPs to discover all PPs in the ecosystem. At the end of a successful fetch-process-validate interval, RPs generate a file with Validated ROA Payloads (VRPs), a list of tuples that map network resources to their ASN owner. Border routers retrieve VRPs via the RPKI-To-Router (RTR) protocol and use VRPs for making routing decisions in BGP. 

{\bf Chain of Trust} in RPKI spans from RIRs down to the end-entities with allocated network resources. All objects in the repositories are cryptographically signed by the host CA's private key, and that CA's resource allocating certificate is provided to the parent CA, who in turn signs it with its own private key. The result is an uninterrupted chain of signed certificates starting with the root RIR certificate. In this way, the RPKI ecosystem allows any RP to validate the integrity and completeness of the objects it receives.

{\bf Objects retrieval and validation}. If an RP is querying a PP for the first time, it will by default do a full snapshot download and save the session\_id and serial\_id. During the follow-up validations, the RP performs a delta update if the session\_id remains unchanged and the RP can bootstrap all the delta increments from its repository cache without errors. If delta fails, the RP falls back to snapshot download. 
 RP validation starts with the TALs of the 5 RIRs, which point to the first repositories in the ecosystem. The RP is obliged to download and process all the valid objects in every repository it lands. Further, it must also jump to every PP it discovers along the way. The validation process can only end after the RP has exhaustively queried all the PPs in the ecosystem, and the final VRP output can only be generated if the RP finishes the process successfully and exits with code 0.
 
{\bf RPKI-To-Router (RTR) Protocol} \cite{RFC6810} is a custom RPKI protocol that interfaces between the RP and BGP routers. RTR is a lightweight protocol that serves the VRP data to routers upon request. It allows both unprotected TCP connections and supports communication through secure channels such IPSec or SSHv2.

\subsection{Key Problems}\label{sc:problems}
Despite the central role that RPKI has in the Internet, there are key problems that limit its effectiveness and its wide adoption. Most of the problems are around the ability of the relying parties to obtain and validate RPKI objects, both due to attacks as well as due to benign failures. We explain them next.

{\bf Connectivity failures.} Each network that wishes to enable ROV filtering for blocking bogus BGP announcements, deploys an RP. The border routers on that network retrieve the VRPs from the RP. If the RP fails at downloading a complete set of ROAs, e.g., due to connectivity failures, once the cached objects become stale, no RPKI validation will be applied. To enhance the availability of RPKI, the networks should deploy a higher number of RPs, so that at least some of them successfully fetch the RPKI objects. Unfortunately, a larger number of RPs increases the overall load on the repositories, which results in higher failures. For instance, we find that repositories limit the number of the RPs that can concurrently connect to the RPKI repository. We find that currently, 4.6\% to 12.5\% of repositories exhibit chronic availability issues ranging from expired TLS certificates to bad formatting of the notification.xml file - the entry point to a repository's data. These issues force the RP client to fall back to rsync, which the deployers have limited service for, namely our measurements indicate that the RPs are rejected because the repository can only serve 50 rsync connections at a time. According to our data collection, there are over 5.6K unique RPs on the Internet competing for access to these repositories with most networks using one RP instance. However, the more RPs the networks deploy, the more connectivity failures the RPs will experience during their attempts to fetch RPKI objects. These types of errors and limitations are expected to grow in the future when RPKI is fully deployed, and multiple networks use RPs to fetch RPKI objects.

{\bf Inconsistencies.} Recent work \cite{ndss:2024} showed that inconsistencies in processing RPKI objects across relying parties are common. Factors for these inconsistencies range from connectivity failures, to different processing logic in each relying party implementation. For instance, Fort implementation of relying party software was found to discard 6,405 VRPs issued by Amazon due to a strict certificate validation logic. The cause was that when the certificates use the option OrganisationName attribute field instead of CommonName or SerialNumber, the RPKI repository was discarded. Consequently, all these prefixes lost the security guarantees of RPKI despite having issued the covering ROAs. Additionally, the networks that were applying ROV with Fort software package not only risked accepting hijacked BGP announcements but could also lose connectivity to those prefixes for RPKI deployments that were in "hard fail mode" and filtered announcements for which valid ROAs could not be retrieved. When the relying party software packages reach different conclusions regarding validity of RPKI objects this also has implications for routing convergence and stability. 

{\bf Attacks exploiting software vulnerabilities.} There has been comprehensive research and studies on the vulnerabilities of relying party software and the RPKI protocol itself. When repositories in the ecosystem host malformed objects, the processing errors often lead to a full crash of the relying party \cite{ndss:2024}. Relying parties require the global validation to finish successfully before it generates the updated VRP output. Under these circumstances, a DoS-triggering repository would lock any relying party in a perpetual state of reboot \& crash, thus never generating a VRP output. These errors can effectively knock out RPs, such that the connection between RP and router is broken and the router stops receiving VRP data. 

{\bf Attacks exploiting protocol vulnerabilities.} Another major attack vector against RPs are stalling attacks \cite{stalloris, hlavacek2023beyond}. These attacks rely on maximizing the connection time between the RP and the repository for a potentially unlimited number of child nodes. Stalling attacks exploit the inflexibility and naiveté of the standardized validation process in RPs, by extending the validation interval long enough for important short-lived objects in the cache, such as manifests, to expire. The expiration of manifests leads to loss of validation for all objects in that repository. Similarly to DoS, stalling attacks lead to VRP loss and thus targeted RPKI downgrade.

Our proposal \brpki\ enhances the resilience of the relying party functionality by distributing it across a number of relying party instances and using Byzantine consensus to ensure consistency of the produced VRPs. Additionally, there are also two side effects to our proposal. First, as more routers use our intermediate relying party functionality, our proposal will in the long term reduce the volume of traffic to the repositories and hence also failures, since the routers will be able to query our central cache directly, instead of individual relying parties querying the repositories separately. Second, since we remove the need for networks to set up and configure relying parties themselves, we expect that our proposal will also facilitate wider adoption of RPKI filtering with ROV.

\section{Design Challenges}\label{sc:challenge}
The design of a system that synchronizes RPKI data from multiple sources, requires the consideration of several RPKI-specific characteristics. To model RPKI object synchronization there is a set of RPKI operational constraints that need to be accounted for. In this section, we describe our observations of the nature of RPKI data and of PP service quality, the challenges that they introduce to the development of a Byzantine relying party, and our design choices to address them.

{\bf Objects interdependencies.} RPKI objects have complex interdependencies between multiple data types. These interdependencies affect the inter-PP and inter-object relationships. Namely, objects within a CA can reference one another, and certificates reference the location of the child CAs, which can be located either within the PP or delegated to another one. Then there is the manifest file, which lists a hash of all objects in the repository and only the files correctly listed in the manifest must be validated according to the standard. If any object is altered without changing the corresponding hash entry in the manifest, or the manifest is updated without matching its contents with existing objects, then the manifest will no longer mirror the correct state of the repository, and valid objects may no longer be correctly validated.

{\bf Cryptographic dependencies.} All RPKI objects are cryptographically signed by the CA's own certificate. CAs enter the RPKI tree by storing their certificate into the repository of the parent CA. The resources of a child CA can be served by the same PP or an externally delegated one. If the parent CA does not contain the most recent certificate for the child delegation, the child CA resources may not be discovered. Additionally, if a certificate is updated the objects of the affected CA need to be updated as well with new signatures. Failure to have an up-to-date chain of signatures and signed objects, results in loss of object correctness during RP validation, and therefore loss of VRPs. 

\begin{table*}[t!]
\scriptsize
\renewcommand{\arraystretch}{0.7}
\centering
\begin{tabular}{l|l l l|l l l}
 \textbf{Experiment} & \multicolumn{3}{c}{\bf no cache}  &  \multicolumn{3}{c}{ \bf with cache}\\
 & {\bf M1} & {\bf M2} & {\bf M3} & {\bf M1} & {\bf M2} & {\bf M3} \\\hline
\textbf{Network A (Host 1)} &406,805&518,233 & 518,537& 518,942 & 521,304 & 521,325 \\\hline
\textbf{Network A (Host 2)} & 406,905 & 518,058 & 513,368& 513,824 & 513,833 & 514,032 \\\hline
 \textbf{Network B} & 407,153 & 517,323 & 507,975 & 514,001 & 514,013 & 514,034 \\\hline
\end{tabular}
\caption{{\small VRP datasets for different setup of RPs.}}
\label{tab:rpki_database}
\end{table*}

{\bf VRPs change frequently and dynamically.} VRP sets are not guaranteed to be the same over an arbitrary period of time. Our local tests have shown that recurring validation processes spread over 10-minute intervals can result in differing sets of VRP results. These issues are prevalent even when the RPs are operating with a valid fallback cache. To verify these observations, we run an experiment consisting of three distributed RPs running the rpki-client software package. We use two different networks for our tests. On one of the networks we pick two different hosts to run our experiments on. Table \ref{tab:rpki_database} shows the discrepancies of the VRP sets, while all 3 distinct RPs run the same software package and were asynchronously instantiated on different networks and hosts. We run the experiments twice: once with empty caches for every iteration, and once with the previous validation cache to fall back on. The validation runs on the different RPs are initiated at \~10 minutes shifts from one another, to emulate the behavior of a distributed asynchronous polling system.

The first thing we observe during {\it no cache} validation runs, is large oscillations of the VRP datasets, both within the same RP instance and different RPs. As a matter of fact, we notice a drop of over 100,000 VRPs and a few minutes later the number goes back up to the expected values. Even when the error that caused the drop is fixed, we still have VRP data jittering by the hundreds for our different RP viewpoints. During the second round of experiments {\it with cache}, we still observe noticeable oscillations of the VRP data over the different hosts, despite our clients having a robust recently populated cache to fall back on. These oscillations within the same querying network, and across different networks, despite an up-to-date fallback cache, suggests that the nature of problems that lead to data jittering range from network level and connectivity issues, to internal inconsistencies in the PP repositories themselves, where any changes such as ROA additions, manifest re-issuance, or key rolling (certificate re-issuances), need time to propagate throughout the database and change all the affected objects. Our observations show that RPKI object caches are often in flux.

{\bf VRP-based synchronization.} After taking into consideration the tenuous inter-repository and inter-object dependencies, and observing via active measurements volatility in the repository services and availability, we decide to develop our peering and synchronization infrastructure around the VRP output layer and not the RPKI object cache. A change in the VRP output is dependent on the cache changes, however, building a synchronization system around these objects, considering the randomly occurring amount of data jittering, could realistically incur the creation of a chimera cache that results in broken dependencies and introduction of new errors, thus causing artificial VRP loss. Individual VRP entries, on the other hand, are independent of each other and can easily be handled via set operations. Under this paradigm, our system cannot introduce new errors by breaking dependencies: the worst-case performance of our synchronization system mirrors the worst-case performance of a live repository. Furthermore, synchronizing VRPs requires far less data transmission, as these are much smaller than the chain of cryptographic objects they are derived from.

\section{Byzantine Consensus}\label{sc:byzantine}
Our design leverages ideas from distributed computing theory research in the presence of Byzantine adversaries. In this section we provide an overview of Byzantine agreement protocol and explain the challenges in applying Byzantine agreement to synchronize the views of the relying parties.

{\bf Byzantine agreement guarantees.} In traditional Byzantine agreement, the goal is for participants to agree on their values. This is accomplished by replicating the service across the group of participants. We use a consensus algorithm to achieve agreement between the distributed instances of the relying parties. Byzantine fault tolerance protocols require a fix set of participants that is determined prior to protocol execution. This ensures security against Sybil attacks. Byzantine agreement protocols work well on a small number of participants and achieve the following guarantees assuming a 2/3 honest majority: \circled{1} {\em Correctness.} Assuming all messages from the benign participants propagate within a bounded time, the agreement can be reached within a constant time. \circled{2} {\em Consistency.} Assuming a Byzantine adversary, all honest participants finish the execution with the same output. 

{\bf Selecting a Byzantine protocol.} To produce a single system-wide output, all nodes in our proposed \brpki\ need to arrive at the same global state. Distributed systems that assume only benign failures typically use state-machine replication algorithms, such as PAXOS or RAFT \cite{paxos,raft}, to ensure that all nodes synchronize to the same, ordered output, regardless of transient failures. However, these algorithms do not provide byzantine fault-tolerance (BFT), i.e., they fail in the face of malicious behavior, and thus are insufficient for our desired security guarantees. In addition, rather than simply \textit{replicating} the state across the nodes, we explicitly want to maintain the redundancy of having each node process RPKI objects to protect against transient failures on the side of PPs.

We also opt against classic BFT algorithms, as they are not well-adapted to our specific use-case. For example, Federated BFT relies on deriving network-wide consensus from overlapping sets of locally trusted nodes. This is not applicable in our scenario because all nodes in our system have already been permissioned to participate, thus we do not need to discriminate to utilize them. 

Other widely-used BFT protocols are HoneyBadgerBFT \cite{honeybadgerbft} and Alea-BFT \cite{alea-bft}, which are focused on the needs of blockchains. Our system requirements do not include an auditable, immutable history, we only need the current state, thus can ignore the overhead and complexity of organizing our updates in blockchain transactions. Unlike blockchain, our system can rely on point-to-point communication, simplifying many of the communication mechanisms, such as atomic broadcasts, of protocols designed for networks that are not fully connected. 

Among the protocols considered, PBFT \cite{pbft} is the closest to a state-machine replication algorithm tolerant to byzantine faults. However, since each validation run of an RP produces a new complete output, each update of the local state of an RP is idempotent, i.e., independent of the previous state. The same goes for the global state, which can be assembled at any given time from the current local states. As such, there is no requirement for ordering of updates in our system, allowing for a simpler consensus algorithm. Finally, similarly to PAXOS and RAFT, PBFT makes the assumption that a given operation executed by a node is deterministic with respect to the result. However, running two RPs within even a short delay may result in slightly different outputs, due to objects added or removed from the PPs. Moreover, we do not want to assume that RPs run at the same time, breaking PBFT's assumption of synchronous communication. We therefore opt for a simple threshold vote.

\section{Byzantine Relying Party}\label{sc:brp}

In this section we introduce the architecture and components of \brpki: an intermediate layer of relying parties that run a consensus algorithm among them, based on the principles of distributed computing research \cite{dolev1986reaching}. \brpki\  consists of a network of peering nodes acting as autonomous entities that synchronize their output to reach an optimal consensus.

\subsection{Node Data}
\label{sec:lists}

Each node in our system stores and maintains 3 types of data: the peerlist, skiplist and VRPs. Both the skiplists and VRPs of each node are used to produce the master skiplist and master VRPs via the threshold vote (see Section \ref{sec:vote}).

\textbf{Peerlist.} Each container is initialized with a list of IP addresses corresponding to existing peers, in order to bootstrap the peering process. This list is continuously updated by the peering component (see Section \ref{sec:components}). Example:

{\footnotesize{
\begin{verbatim}
    172.17.0.2
    172.17.0.3
    ...
\end{verbatim}
}}

\textbf{Skiplist.} We use the open-source rpki-client as our RP implementation. A key advantage of rpki-client is its integrated support for blacklisting problematic or malicious PPs that should be ignored during a validation run via the "skiplist". The skiplist is maintained by the monitoring component (see Section \ref{sec:components}) and contains a list of PP domains that were detected as problematic, i.e. either stalling the RP or causing it to crash. Via the threshold vote, the nodes produce a master skiplist, which is ultimately used by rpki-client to skip the offending PPs in subsequent updates. Example:

{\footnotesize{
\begin{verbatim}
    rpki.ripe.net
    rrdp.arin.net
    ...
\end{verbatim}
}}

\textbf{VRPs.} The VRPs are produced by the RP and stored in JSON format. For each type of RPKI data (currently ROAs, ASPAs and BGPSEC keys), the JSON file contains a list of unique items, representable as strings, encoding each VRP. Each of these sublists is aggregated across the nodes via the threshold vote to produce the JSON file containing the master VRPs. This master output is served to routers via RTR. Example:

{\footnotesize{
\begin{verbatim}
{ 
    "roas": [
    {"asn":"AS65537","prefix":"192.0.2.0/24",
     "maxLength":24,"ta":"ARIN"},
    {"asn":"AS12345","prefix":"2001:db8::/32",
     "maxLength":48,"ta":"RIPE"},
    ...
  ],
...}

\end{verbatim}
}}

\subsection{Node Components}
\label{sec:components}

\begin{figure}
    \centering
    \includegraphics[width=0.75\columnwidth]{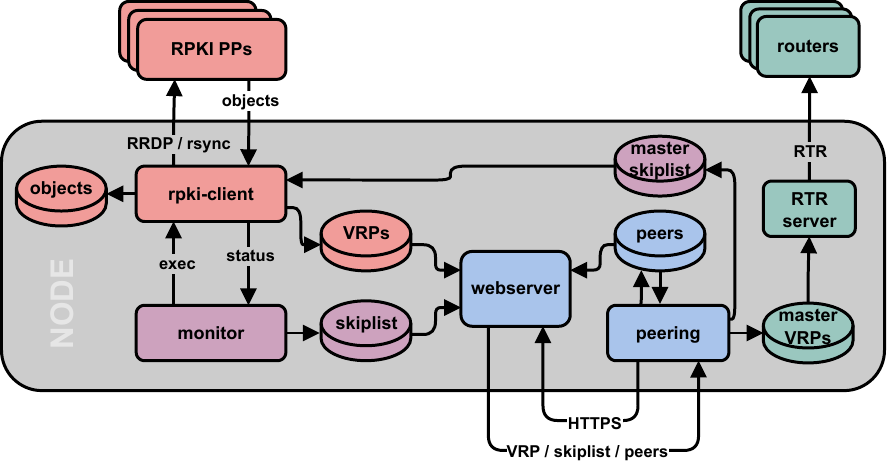}
    \caption{{\small Container components represented as squares, data/files as cylinders. Color legend: red - standard RPKI; purple - detection/skiplisting; blue - peering/consensus; green - client output.}}
    \label{fig:design}
\end{figure}

\brpki\ is a fully-connected network of nodes, implemented as Docker containers. Each container is a self-sufficient instance containing an RP and an associated monitoring process, a peering management process, an RTR server and a webserver (see Figure \ref{fig:design}).

\textbf{RP (rpki-client).} We use the open-source rpki-client as our RP implementation, which fetches RPKI objects from PPs and validates them to produce the node's (local) VRPs. As described in Section \ref{sec:lists}, rpki-client uses the (master) skiplist to skip over PPs identified as problematic.

\textbf{Monitor.} The monitor performs a supervisory function for the RP. It periodically invokes rpki-client, which operates unaltered, but in order to detect stalling attacks, the monitor also starts a packet sniffer to keep track of how long connections to each PP last. Based on the RP exit code, the monitor can also detect when the RP crashes and on which PP. Both stalling and crashing lead to the addition of the offending PP to the (local) skiplist. We explain how exactly the monitor ensures resilience against adversaries in Section \ref{sc:smart:monitor}. The monitor also stores a timestamp for each skiplist entry and removes entries that have passed a configurable expiration time to avoid skipping PPs indefinitely.

\textbf{Webserver.} The webserver is used to host the files detailed in Section \ref{sec:lists}: the local skiplist produced by the monitor, the local VRPs produced by rpki-client, and the peerlist. The webserver provides a simple method for point-to-point, pull-based communication. All peer communication is protected by mutually authenticated TLS, based on a single trust root certificate. This means peers only communicate with peers that have a valid certificate signed by the trust root. This creates a permissioned network. Each container is initialized with such a certificate for its IP address (or domain name) and can validate peers' certificates using the root certificate. The root certificate itself is also included in the container build, but can be rotated by rebooting and mounting the new certificate into the container.

\textbf{Peering.} The peering process is responsible for both maintaining an up-to-date peerlist and aggregating the skiplists and VRPs. It periodically polls all known peers for their peerlist, local skiplist and local VRPs. The skiplists are aggregated via the {\em threshold vote} (see Section \ref{sec:vote}) to produce the master skiplist, which is fed to rpki-client. The VRPs are aggregated analogously to produce the master VRPs. Starting with an initial list of peers, provided when the container boots, the peerlist is updated by mutual polling. "Candidate" IP addresses are obtained through a) checking the peerlists of known peers for IPs not yet included in the local peerlist, and b) checking the webserver access logs for IPs that requested the local peerlist. All candidate IPs are then queried for their peerlist. If the query succeeds, the IP addresses are added to the local peerlist, ensuring a fully-connected network.

\textbf{RTR server.} Each node also contains an RTR server (in our case, stayRTR), which simply serves the post-consensus master VRPs\footnote{Because stayRTR requires a URL from which to obtain the VRPs, the master VRPs are made available to the localhost via the webserver.} to the border routers via the RTR-over-TLS protocol (details follow).

\subsection{Threshold Vote}
\label{sec:vote}

\begin{figure}[th]
    \centering
    \includegraphics[width=0.4\columnwidth]{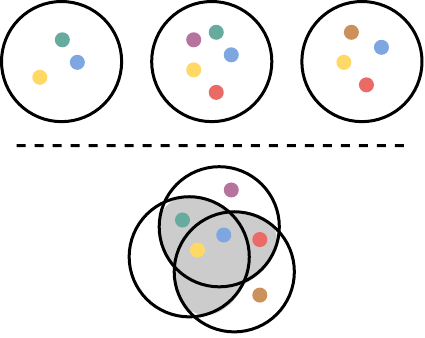}
    \caption{{\small Threshold vote for 3 nodes and resilience to 1 fault ($f=1$).}}
    \label{fig:vote}
\end{figure}

After separately fetching the RPKI objects from the RPKI repositories and producing the VRPs, each peer obtains the VRP set of each other peer and computes the master VRP set, representing the consensus of the network, using a threshold vote:

\textbf{Definition.} For a set of peers $P$, $L_p$ denotes a local set of independent objects (in our case, strings representing VRPs or skiplist entries). The global set of objects, representing consensus among the peers, is computed as $G = \{o \in \bigcup_p^{p \in P} L_p ~|~ v(o) \leq f + 1\}$, where $v(o) = |\{p \in P ~|~ o \in L_p\}|$. This set includes all objects that appear in the local set of at least $f+1$ peers, where $f$ denotes the maximum number of failing nodes (both benign and malicious) that the system can tolerate. Figure \ref{fig:vote} visualizes this for a network of 3 nodes and $f=1$, requiring at least 2 votes for each object.

In our system we compute $f$ as a function of the current number of peers in the network and a factor $c \in [0, 1]$, which allows easy parameterization of the consensus requirement and its dynamic adaptation to peers joining the network: $f = \lfloor c * |P| \rfloor$. A consensus factor of $c = 0$ therefore makes the global set equivalent to the union of all local sets: $G = \bigcup_p^{p \in P} L_p$, whereas a factor of $c = 1$ results in a global set equal to the intersection of all local sets: $G = \bigcap_p^{p \in P} L_p$. In our experiments we use a default value of $c = 0.5$.

\textbf{Properties.} This voting method makes minimal assumptions to function, requiring only authenticated point-to-point communication, so malicious nodes cannot manipulate the content communicated by legitimate nodes or take advantage of network topology to cut nodes off from the network.

Consensus can be formed for each object independently, it also adapts to changing numbers of observed peers, it is agnostic to benign failures (e.g., unreachable peers) or to byzantine behavior (e.g., injection or censoring of objects), and can run asynchronously.

Injecting bogus objects ("poisoning") requires a sufficient fraction of malicious nodes to exceed the consensus threshold ($f + 1$), while censoring an object, that is otherwise supported by all legitimate nodes, requires $|P| - (f + 1) - d$ malicious nodes, where $d$ denotes legitimate nodes failing or disagreeing on the object in question. This applies both to VRPs and skiplist entries.

\subsection{Distributing VRPs to Routers}

The master VRPs produced by the consensus of the networked peers is securely served by each container via RTR-over-TLS. Border routers should therefore download the VRPs directly from the nodes, eliminating the need for users to run their own RP instances. By downloading directly from the nodes, we avoid the use of a dedicated RTR server that would introduce a central point of trust and failure (which would negate the purpose of the consensus network). To facilitate this the service operates under a domain name with DNS type A resource record configured for each peer node IP address. This also provides built-in load balancing of client requests across the nodes.

{\bf Auditing bogus VRPs.} A malicious node can supply arbitrary VRPs to client routers. However, clients (or dedicated monitors) can at any point query all nodes via their IP address and compare their outputs, allowing for transparent auditing of each node.

{\bf Independent verification of VRPs.} Furthermore, the validity of each VRP provided by the service, can be independently verified by users running their own RP and comparing VRPs. Due to the possibility of RPKI objects being removed from a PP or revoked between validation runs of any two RPs, it is not feasible to conclusively refute the \textit{absence} of a given VRP, but the \textit{presence} of a VRP in the service's output can be double-checked via the users' RP output. A fraudulent VRP in the master output would not have the corresponding RPKI objects to back it up. As we explain in Section \ref{sc:smart:monitor}, it would consistently fail to be produced by an independent RP.
Thus, the service as a whole is open and it can be independently verified, which disincentivizes misbehavior by malicious nodes.

\subsection{Monitoring Malicious and Faulty PPs}\label{sc:smart:monitor}
Another aspect of the infrastructure is robustness against malicious or faulty PPs. Malicious PPs can stall validation of RPs or trigger attacks against the RPs using unknown or unpatched vulnerabilities or processing errors. These issues can land the RP in a perpetual state of fail-restart due to the unflexible nature of the RP validation algorithm. This state stops the VRP information flow towards the BGP routers, thus it can downgrade RPKI protection \cite{van2022rpkiller}. Worse, the fail-restart loop that characterize these errors can persist for multiple validation rounds, even days, requiring manual intervention from an operator. We design a set of algorithms that ensure robustness and guarantee correctness and availability of VRP results under any circumstances.

All nodes are equipped with a dedicated monitoring and forensics mechanism that analyzes and logs all TCP connections established by the RP software. Algorithm \ref{alg:conn_algo} models the TCP connections. Every established connection is logged as a dictionary with a start- and end-time. This dictionary allows the system to track which connections have been open for too long and which connections were ongoing if the RP crashes. The corresponding repository's domain is then added to the skiplist.

\begin{algorithm}[ht!]
  \scriptsize
    \KwIn{\textit{$<$stream$>$} packets} 
      \KwOut{\textit{$<$dict$>$} connections\_log} 
 \While{True}{
 \If{syn in OUT\_packet.flags}{
 connections\_log[IP\_dst]["start time"]=now()\\
  connections\_log[IP\_dst]["established"]=False\\
 connections\_log[IP\_dst]["end time"]=None\\
 }
 \If{syn,ack in IN\_packet.flags}{
   connections\_log[IP\_src]["established"]=True\\
 }
 \If{RST, FIN in IN\_packet.flags or OUT\_packet.flags}{
    connections\_log[IP\_src or IP\_dst]["end time"]=now()\\
 }
 }
 \caption{Connection Logging Algorithm}
  \label{alg:conn_algo}
\end{algorithm}

{\bf Adding crash-triggering PP to skiplist.} The monitor operates in accordance to Algorithm \ref{alg:crash_algo}. It continuously polls the state of the RP. If the RP fails for any reason, the monitor checks which connections were still open during the crash, and logs this information into the local skiplist as a potential crash-triggering PP. If the crash manifests in other RPs, a majority of nodes eventually agrees on the connections that lead to the crash of their respective RP. The agreed upon domains are then added to the global master skiplist, which is enforced throughout the cluster.

The domains stay on the master skiplist for a configurable amount of time defined in the global variable {\it blacklist\_expiry}. Blacklisting of repository domains should not be indefinite to account for false positives. For example, sometimes issues are caused by misconfigurations, which are eventually fixed by the operators. After a skiplist entry expires, RPs will resume querying the domain in the next validation round.

\begin{algorithm}[ht!]
\scriptsize
  \KwIn{\textit{$<$dict$>$} dnsbook}
  \KwIn{\textit{$<$dict$>$} connections\_log} 
  \KwOut{\textit{$<$list$>$} anomalous} 
  \If{ RP.crash {\bf is} True}{
   \ForEach{conn {\bf in} connections\_log}{
   \If{conn["end time"] {\bf is} {\it None} {\bf and} conn["established"]}{
    anomalous.append(dnsbook[conn])\\
   }
   }
   {\bf return} anomalous
}
 \caption{Crash Detection Algorithm}
  \label{alg:crash_algo}
\end{algorithm}

{\bf Adding stalling PP to skiplist.} Stalling attacks require a more finely tuned detection approach due to their characteristics, as described in Section \ref{sc:problems}. Such attacks would be undetectable by our crash detection mechanism because the RP neither crashes nor exits.

We design a custom algorithm to monitor for stalling attacks. Our system keeps track of a connection's lifetime throughout the RP validation interval, and continuously monitors the longevity of any connection, see Algorithm \ref{alg:stalling_algo}. When the monitor detects that a connection is taking long enough to reach the configured global timeout of the RP, we disconnect the RP and log the suspect repository domain into our skiplist. By default, rpki-client times out individual connections to repositories at $\frac{1}{4}$ of the global timeout.

Given that the median refresh duration across all repositories is around 2 minutes, a major deviation would be suspicious. Therefore, we treat connections that are getting close to the connection timeout as maliciously attempting to maximize their longevity. With this approach, we can protect our system RPs from getting stuck while crawling malicious nested branches. 

\begin{algorithm}[ht!]
\scriptsize
  \KwIn{\textit{$<$dict$>$} dnsbook}
  \KwIn{\textit{$<$dict$>$} connections\_log} 
  \KwIn{\textit{$<$datetime$>$} start}
  \KwIn{\textit{$<$parameter$>$} threshold}
  \KwOut{\textit{$<$list$>$} anomalous} 
  \ForEach{conn {\bf in} connections\_log}{
   \If{conn["end time"] {\bf is} {\it None} {\bf and} conn["established"]}{
    \If{conn.length $>$ threshold*timeout}{
     anomalous.append(dnsbook[conn])
    }
  }
  }
   {\bf return} anomalous
 \caption{Stalling Detection Algorithm}
  \label{alg:stalling_algo}
\end{algorithm}

{\bf Monitoring for attacks.} After putting all the above-mentioned subroutines together, we design the monitoring Algorithm \ref{alg:monitor_algo}. While the connections are logged in a background process, our monitor continuously polls the state of the RP to detect if any crashes are present. If there are crashes, the skiplist is updated and the validation cycle continues with the next TAL in the queue. If there are no failures and the RP is still running, the system continuously monitors the length of ongoing connections to detect any stalling attacks or abnormally long connections. If any are detected, the RP is killed, the skiplist is updated, and validation continues with the next TAL.

\begin{algorithm}[ht!]
\scriptsize
  \While{True}{
  \ForEach{\upshape tal \textbf{in} TALS}{
   start relying\_party \\
   \While{True}{
    current\_time $<-$ now()\\
    skiprepos $<-$ stalling\_detection(current\_time)\\
    update\_skiplist(skiprepos)\\

    status $<-$ relying\_party.poll()\\
    \If{status is not None}{
     \If{status is 0}{
      aggregate\_vrps\_to\_master()\\
      break \\
     }
     skiprepos $<-$ crash\_detection()\\
     update\_skiplist(skiprepos)\\
     break \\
    }
    sleep \\
   }
   }
   aggregate\_vrps\_to\_master()\\
   sleep() \\
   shuffle\_tals() \\
  }
 \caption{Monitor Algorithm}
  \label{alg:monitor_algo}
\end{algorithm}

\subsection{Validation Rounds Required to Converge} 

Generally, convergence requirements under RPKI networks depend heavily on a number of factors: \circled{1} network conditions of the PP \circled{2} availability of the PP to accommodate further concurrent connections \circled{3} backend synchronization of objects. These are parameters beyond the control of an RP.

Convergence can be reached once a majority of nodes fetch and validate objects from all 5 TALs. Timewise, this depends on networking conditions necessary to finish the downloading and validations of the global RPKI data. Currently, the average time to download and validate the entire global RPKI dataset for the first time is around 3 to 5 minutes for rpki-client in a well-connected network. Our peering infrastructure polls for updates every 10 seconds, thus convergence can be achieved with little delay as soon as a majority of nodes have finished downloading the first set of global RPKI data. Each RP validation round works one TAL at a time, and all RPs operate on a randomized set of TALs, in order to maximize download diversity and speed. Instead of running all TALs simultaneously, we split them, and then we put the results together into the final VRP file once all TALs per RP are processed. Furthermore, we ensure that during the first run of all nodes, they all start on a different TAL, and then the following TALs are randomized from the leftover list. This allows our system, to diversify node results.

\section{Security Analysis}\label{sc:security}

In this section, we define adversarial capabilities and explain how our system handles errors and attack vectors as listed in Section \ref{sc:problems}.
The three core design decisions of our proposed system are \textbf{a)} individual validation of TALs, \textbf{b)} redundancy through the use of multiple RPs with a single, consensus-driven output, and \textbf{c)} adaptive blacklisting of malicious PPs. All three contribute to higher availability of VRPs and thus better security, as discussed in Section \ref{sec:repo-attacks}. However, the consensus aspect introduces new attack vectors, which are discussed in Sections \ref{sec:poisoning} and \ref{sec:inconsistency}.

\subsection{Threat Model}

We consider an adversary that can corrupt or block a limited number of RPs in our system. Additionally, this adversary can also host PPs with arbitrary content to stall or crash RPs.

Following the principles of the Byzantine agreement protocol,
we assume that the majority of RPs are honest and not corrupted. An adversary that corrupts a majority of RPs can trivially manipulate consensus. Furthermore, a MitM between the RPs and dependent routers or between the RPs and PPs is out of the scope of this paper, since it can arbitrarily drop messages and trivially break the security of RPKI by blocking availability of VRPs or RPKI data.

\subsection{Manipulating VRPs}
\label{sec:poisoning}

\brpki\ is robust to poisoning attacks and preserves correctness.

\textbf{Poisoning.} As detailed in Section \ref{sec:vote}, forcing the network as a whole to adopt an arbitrary VRP requires $C = f + 1 = \lfloor c * |P| \rfloor + 1$ malicious nodes to collude, where $c$ is the consensus factor (e.g., 0.5) and $P$ denotes the full set of nodes in the network. This is because an arbitrary VRP would not be derived from RPKI data by legitimate nodes and all votes required to meet the threshold would need to come from the malicious nodes.

\textbf{Censoring.} Conversely, the requirement for censoring a VRP is the inverse, i.e. enough malicious nodes to prevent the legitimate nodes from meeting the threshold: $|P| - C - d$ where $d$ denotes a minority of legitimate nodes that are in disagreement. In transitionary periods, where only part of the network has updated a given VRP, censorship will be easier to achieve, because $d\neq0$ (see Section \ref{sc:challenge} for further discussion). However, the expectation is that all nodes eventually update, leading to $d=0$. Censorship thus only \textit{delays} consensus.

\textbf{Correctness.} The output of the system reflects the outputs of the majority of nodes. The output of each node is equal to the output of the RP. Thus, if the RP functions correctly and the specified numbers of malicious nodes are not exceeded, the output of the system will be correct.

\subsection{Inducing Inconsistency in Consensus}
\label{sec:inconsistency}

Due to the threshold vote, malicious nodes can produce inconsistencies in the outputs of legitimate nodes. This is possible when a given object (VRP or skiplist entry) has $C - f$ votes, allowing the malicious nodes to selectively push the votes over the consensus threshold for some nodes, but not others, by sending different messages to different nodes.

Such an inconsistent state could easily be \textit{detected} by simply querying and comparing the master outputs produced by each node. Furthermore, as in more general, classic BFT protocols, the malicious nodes could be \textit{identified} through gossiping mechanisms, provided $f \leq \frac{n-1}{3}$. Each node can simply tell all other nodes what it was told from each other node, allowing nodes to compare notes.
However, this requires nodes to work synchronously. Otherwise if, e.g., nodes $A$ and $B$ receive different answers from node $C$, because $C$ updated its output between queries, $C$ would be falsely perceived as malicious. Regardless, we choose to maintain asynchrony, for two reasons: \textbf{a)} avoiding added complexity required to coordinate synchronization between nodes, thus allowing us to maintain minimal operational assumptions, and \textbf{b)} safeguarding against more fine-tuned stalling attacks \cite{hlavacek2023beyond}, due to randomization of the order and relative timing of nodes' individual validation runs.

We argue that inducing inconsistency as outline above does not afford the attacker any advantage, because \textbf{a)} inconsistent outputs would only be transient, since all legitimate nodes would eventually update to the same state for any given VRP and the output of the system would converge, and \textbf{b)} malicious nodes could only cause inconsistency opportunistically for whatever VRPs are currently in flux across the network.
We thus do not consider such manipulation critical to the security of the system.

\subsection{Repository-based Attacks}
\label{sec:repo-attacks}

As shown in previous work \cite{stalloris, hlavacek2023beyond, ndss:2024}, malicious PPs can stall RPs to timeout or crash them with maliciously crafted objects, resulting in missing RPKI data and ultimately unavailability of VRPs. Due to the introduction of a supervisory process (see "monitor" component in Section \ref{sec:components}) in our design, nodes can detect such attacks and add offending PPs to a skiplist. As demonstrated in Section \ref{sec:eval-dos}, successful skiplisting leads to the avoidance of these PPs and the exclusion of content from these PPs, thus the RPs can continue validation uninhibited.

Furthermore, because TALs are validated individually and their order is randomized in each node, stalling attacks have a much smaller impact on the overall availability of RPKI objects. While a given RP is stalled on a TAL with a malicious PP, other RPs will be validating other TALs, unaffected, and contributing the resulting VRPs to the system's master output. This results in higher availability and thus better security.

\begin{figure*}[t!]
    \centering
    \includegraphics[width=0.32\textwidth]{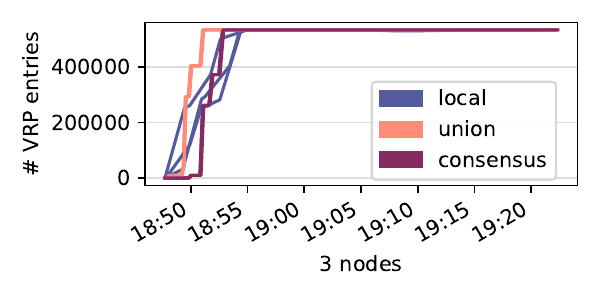}
    \includegraphics[width=0.32\textwidth]{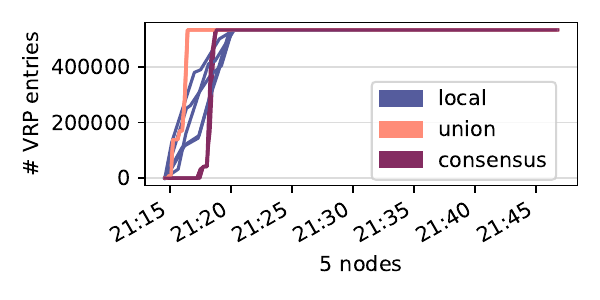}
    \includegraphics[width=0.32\textwidth]{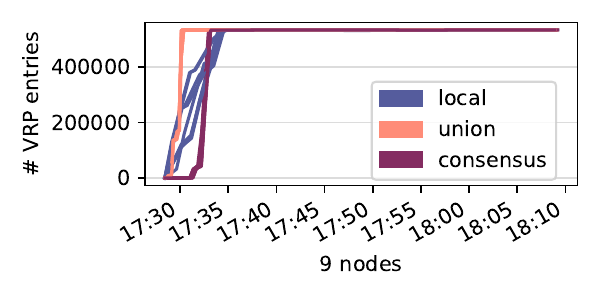}
    \includegraphics[width=0.32\textwidth]{img/benign_9_TE_3600_INT_60.pdf}
    \includegraphics[width=0.32\textwidth]{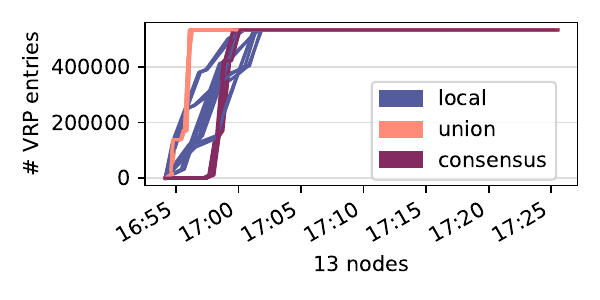}
    \includegraphics[width=0.32\textwidth]{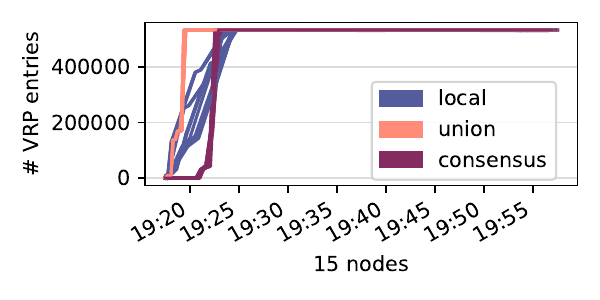}   
    \caption{{\small Experiment A: convergence graph for a benign cluster, 10 seconds refresh interval.}}
    \label{fig:60b}
\end{figure*}

\section{Evaluations}\label{sc:eval}

In this section, we evaluate the speed, convergence and correctness of our system against a range of parameters. For testing, we deploy multiple containers on a VM running Ubuntu 22.04 with 2TB of RAM and 40 cores. This is intentionally chosen to leave a large headroom compared to the resource usage of the containers (see Figure \ref{fig:resource}) to avoid potential resource conflicts. We note that because we deploy the containers on a single machine in a virtual network (Docker bridge), inter-peer network latency is not represented in the convergence time. We evaluate \brpki\ on the real RPKI, interacting with public repositories, and measure its performance under benign and malicious conditions.

\subsection{Evaluations under Benign Conditions}

\subsubsection{Consensus Convergence}

We run our clusters under real network conditions exposed to the global public RPKI repositories. We deploy \brpki\ in cluster sizes ranging from 3 to 15 nodes and let them run for 30 minutes each. The staleness tolerance is set to 1h, the peering polling frequency is set to 10 seconds, and the refresh intervals for the RPs are set to 10 seconds for Experiment A and 10 minutes for Experiment B. 

\begin{figure*}[t!]
    \centering
    \includegraphics[width=0.32\textwidth]{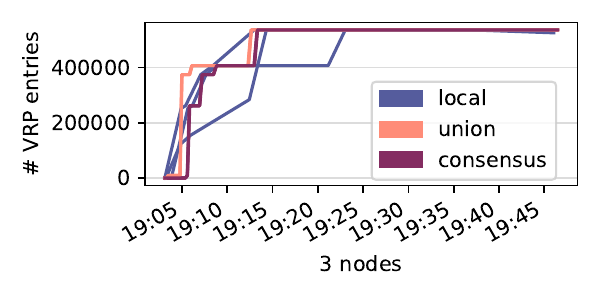}
    \includegraphics[width=0.32\textwidth]{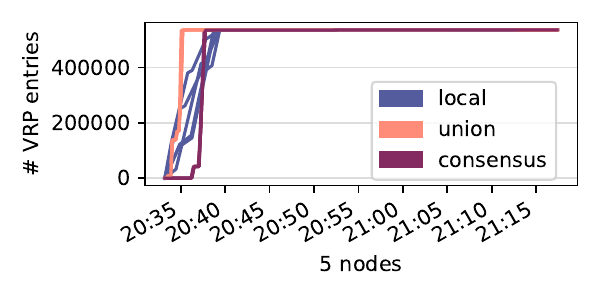}
    \includegraphics[width=0.32\textwidth]{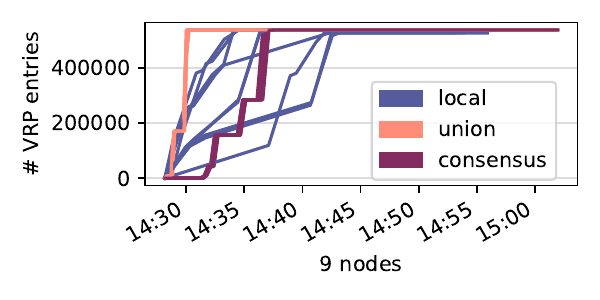}
    \includegraphics[width=0.32\textwidth]{img/benign_9_TE_3600_INT_600.pdf}
    \includegraphics[width=0.32\textwidth]{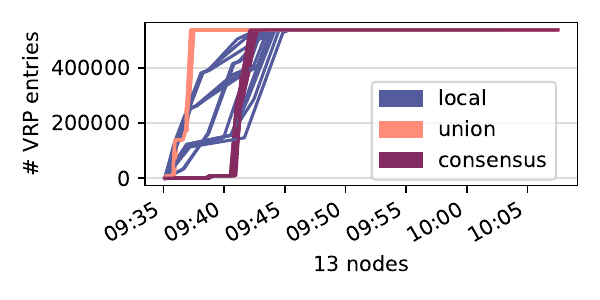}
    \includegraphics[width=0.32\textwidth]{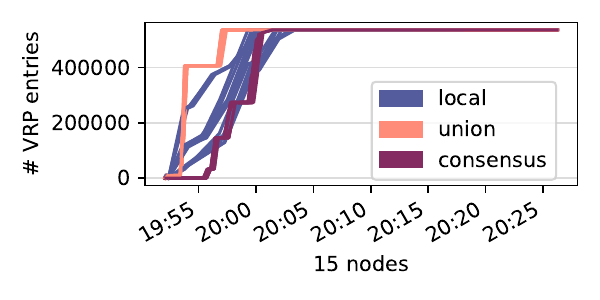}   
    \caption{{\small Experiment B: convergence graph for a benign cluster, 10 minute refresh interval.}}
    \label{fig:600b}
\end{figure*}

Figures \ref{fig:60b} and \ref{fig:600b} show the convergence of our clusters under Experiment A and B. In these graphs, {\it local} is the total number of VRPs aggregated by a single node, {\it union} is the total aggregate of VRPs as detected by all nodes, {\it consensus} is the number of VRPs voted on by a majority of nodes. We observe the master VRP dataset under benign conditions for each cluster size, ranging from 3 to 15 nodes. Our system takes consistently less than 5 minutes for Experiment A and 5 to 10 minutes for Experiment B to reach stable converge in the consensus graph. Clients are served the results from the converged consesus dataset. The {\it local} lines for the individual nodes show that initially not all nodes converge at the same speed, this is due to network performance and service availability of RPKI repositories, but ultimately once the majority of nodes finish full validation, the consensus is immediately established.

\subsubsection{24h Evaluation of correctness and resilience}
To test the robustness and correctness of our system over an extensive period of time, we deploy a 5 node cluster for 24h on the Internet and configure it to use the global public RPKI repositories. Simultaneously, we start a standalone RP instance on the same machine, that refreshes at regular intervals. We measure the VRP output of both instances to compare the efficiency of our system against standard RP deployments.

\begin{figure}[t!]
  \centering
  \begin{minipage}[t]{0.23\textwidth}
    \centering
    \includegraphics[width=0.99\textwidth]{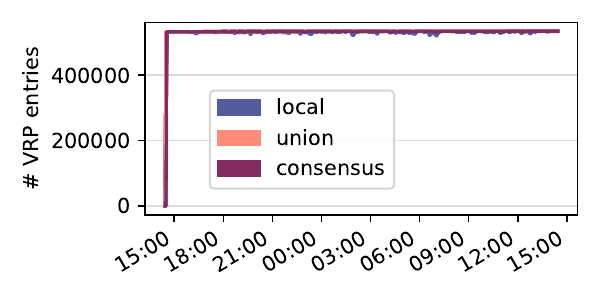}
    \includegraphics[width=0.99\textwidth]{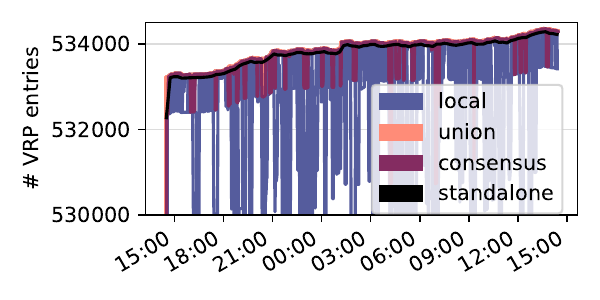}
    \caption{{\small 24h-Experiment without cache: (a) VRP datasets of our system (b) overlap of VRP datasets of our system and rpki-client.}}
    \label{fig:24ben_online_nocache}
  \end{minipage}
  \hfill
  \begin{minipage}[t]{0.23\textwidth}
    \includegraphics[width=0.99\textwidth]{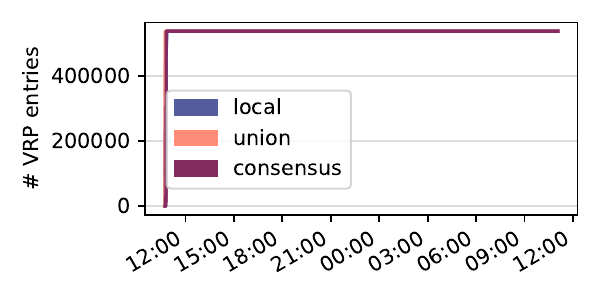}
    \includegraphics[width=0.99\textwidth]{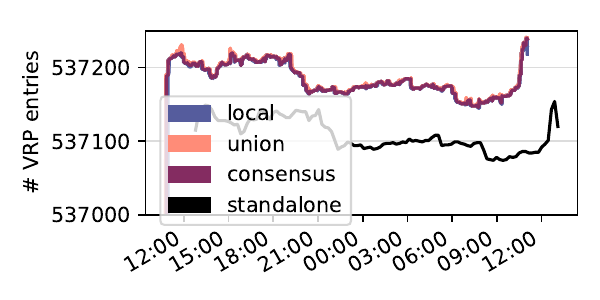}
   \caption{{\small 24h-Experiment with cache: (a) VRP datasets of our system (b) overlap of VRP datasets of our system and rpki-client.}}
   \label{fig:24ben_online_cache}
  \end{minipage}
\vspace{-10pt}
\end{figure}

{\bf Correctness.} We run two experiments with different caching logic. In Figure \ref{fig:24ben_online_nocache} we deploy our cluster without any caching from previous runs i.e., everytime the RP goes through a validation round, the cache is reset before the next round. On the other hand, in Figure \ref{fig:24ben_online_cache}, we persist the caches from the previous run, so every new validation run has the old cache to fallback on in case of networking issues. Figure \ref{fig:24ben_online_nocache} shows that our system remains robust over a 24h run, however, the occasional small {\it local} line dips visible in graph (a) tell us that during some validation rounds, our nodes lose visibility of some data. In Figure \ref{fig:24ben_online_nocache} (b) we see a closeup of our system's VRP collection size and the standalone RP. We can clearly detect frequent major dips in the amount of VRPs that are visible for our system nodes. This is accompanied by an appropriate decrease in the number of VRPs in the consensus line. Comparatively, the standard deviation of the VRP dataset during this 24h monitoring was 340 for the standalone RP client, and 500 for our system. 

We suspect the inconsistency is due to networking issues and objects not being consistently downloaded, so in the next experiment we introduce permanent caching for our nodes, i.e. every validation interval does not start with an empty cache but has the cache of the previous run available. Figure \ref{fig:24ben_online_cache} (a) shows that our system is still robust over 24h, and we observe no {\it local} line dips in the graph, like with Figure \ref{fig:24ben_online_nocache} (a). In Figure \ref{fig:24ben_online_cache} (b) we provide a close up of the graph line and we clearly that there are no longer abnormally frequent and large VRP losses from our system's point of view. The VRP line follows a similar trend to the standalone RP line. This means our system has the same data visibility as any other standard RP. The standard deviation of VRPs for both our system and the standalone client is at 21. Our system does not introduce VRP errors.

{\bf Resilience.} Over the course of our longterm experiments, we encounter circumstances where the RPKI network introduces occasional large VRP losses to querying RPs due to networking or connectivity issues. Figure \ref{fig:24ben_error_wcache} shows side-by-side the performance of our system and the standalone RP when such blackouts occur. We see that RPs are sensible to short-lived blackouts and can immediately disseminate anomalous behaviors. On the other hand, our system is robust against short-lived errors. This is due to the distributed and randomized order in which our nodes query the PPs. The requirement for all nodes to query at different points in time and soon after reach a majority consensus, protects our system from sensitivity to sporadic errors. As a result, our system does not propagate anomalous behavior and has a robust output.

\subsection{Evaluation under Adversarial Conditions}
\label{sec:eval-dos}

In this section, we evaluate the behavior of our system under attack. We simulate a DoS downgrade that is triggered as soon as each node's RP lands on a specific PP. Our attack does not affect the operation of PPs on the Internet, we only artificially cause a crash in our cluster. We do this by inserting a timer inside the container code, that crashes the RP in the node as soon as it reaches the target repository. Our attack simulation targets the RIPE repository and forces all nodes to crash as soon as they land on the target repository. We allow our cluster to operate normally for the first 15 minutes of activity and then trigger the crashes. 

\begin{figure}[t!]
  \centering
  \begin{minipage}[t]{0.23\textwidth}
    \centering
    \includegraphics[width=0.98\textwidth]{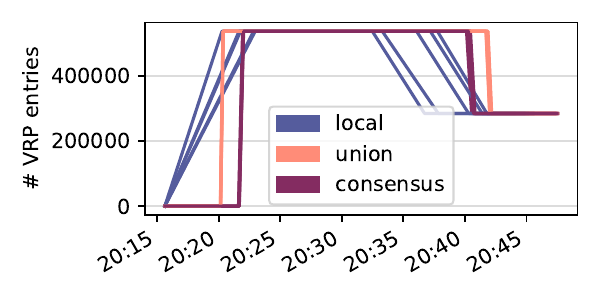}
    \caption{{\small DoS on RIPE TAL.}}
    \label{fig:dos5}
  \end{minipage}
  \hfill
  \begin{minipage}[t]{0.23\textwidth}
   \includegraphics[width=0.98\textwidth]{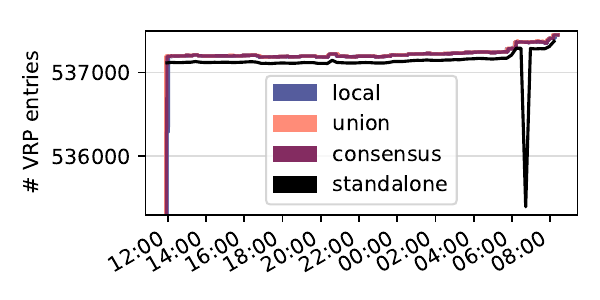}
    \caption{{\small Resilience to errors.}}
    \label{fig:24ben_error_wcache}
  \end{minipage}
\vspace{-10pt}
\end{figure}

Figure \ref{fig:dos5} shows a 5 node cluster under a DoS downgrade attack. We see the first drop in the {\it local} graph lines precisely 20:30, 15 minutes into operations, which means the RP that was currently validating the RIPE repository crashed as soon as it established a connection. However, a single crash does not influence the entire infrastructure. Due to all RPs operating on a randomized order of TALs, it takes a few more minutes for other RPs to hit the RIPE repository as well, and crash. As we see, over the next 10 minutes, multiple other nodes are showing a dip in their {\it local} graph and ultimately, after 3 nodes (the majority in a 5 node cluster), we see the {\it consensus} graph is finally affected and drops to a new plateau without the RIPE repository VRP data. Under current ordinary RPKI setups, regardless of which RP is deployed, such a crash would cause the entire RP to go down and get stuck in a perpetual state of fail-and-restart, never finishing a validation round. As a result, routers would no longer receive updates, thereby downgrading RPKI protection for the entire RPKI ecosystem. This issue demands that an operator jumps in to manually bypass the problematic PP. As we can see in our graph, our system remains up and running, the VRP collection and serving unaffected, despite most RPs crashing on the malfunctioning repository. Our skiplist prevents our RPs from contacting this repository for a period of time, but nonetheless the system will not lose visibility of the VRP data from the other unaffected repositories. Our system is self-healing.

\subsection{Cost of Operation}

We deploy our system under a range of different setups, and measure resource consumption. We run our system in multiple 30 minute intervals, at increasing cluster sizes, starting with 3 and ending with 15. Table \ref{tab:cost} shows the resource consumption of our system for different cluster sizes as reported by \texttt{docker stats} in 10 second intervals.

\begin{figure}[t!]
  \centering
  \begin{minipage}[t]{0.236\textwidth}
    \centering
    \includegraphics[width=0.99\textwidth]{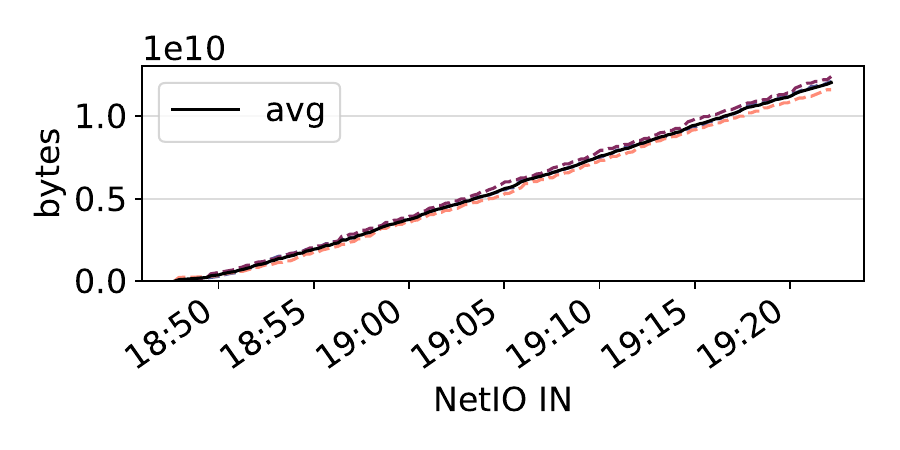}
    \includegraphics[width=0.99\textwidth]{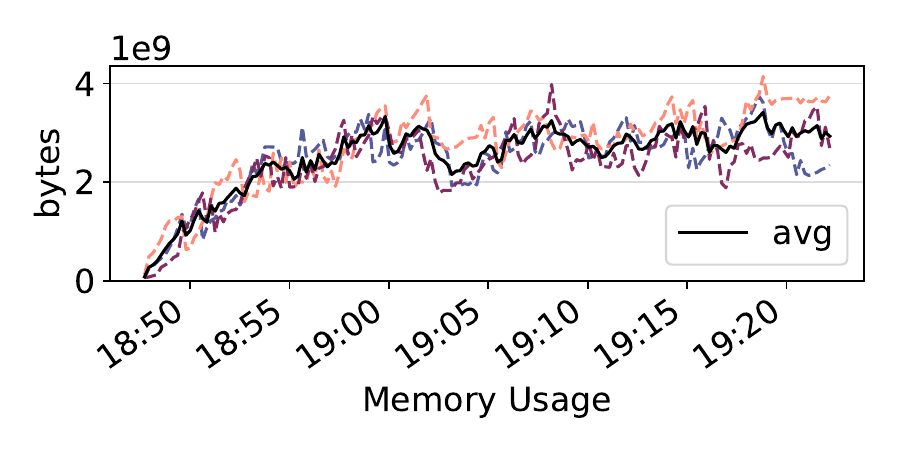}
  \end{minipage}
  \hfill
  \begin{minipage}[t]{0.236\textwidth}
 \includegraphics[width=0.99\textwidth]{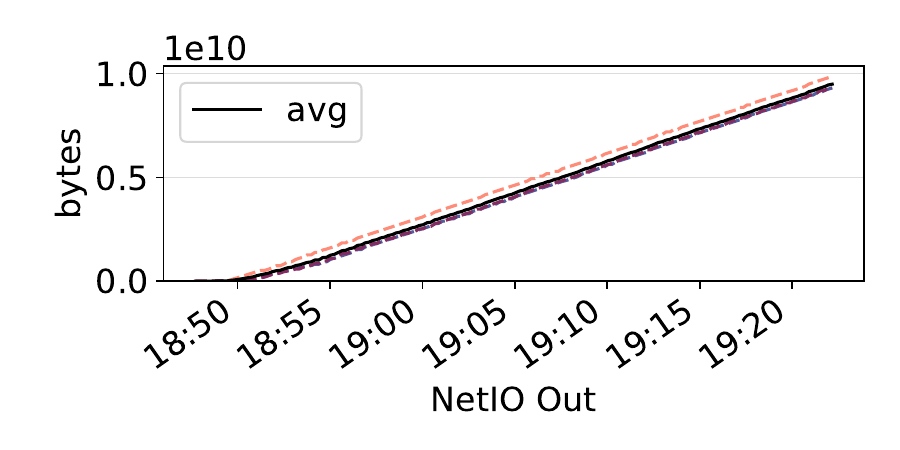}
  \includegraphics[width=0.99\textwidth]{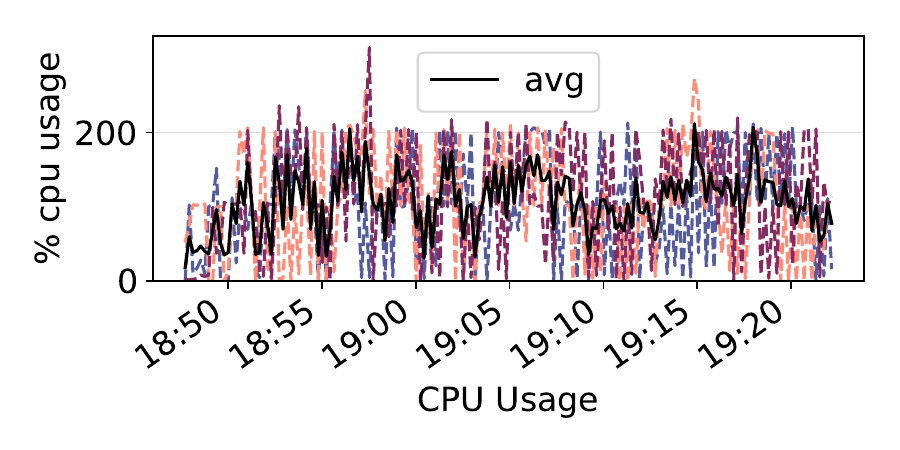}
  \end{minipage}
  \caption{{\small Resource consumption per container for 3 node cluster.}}
\label{fig:resource}

\end{figure}

{\it NetIO} gives the aggregate number of bytes containers receive (IN) and send (OUT) over their network interface. Note that peer communication is counted both as IN and OUT, showing that a large portion of network communication comes from the downloading of VRPs between nodes (which is done every 30 seconds). However, this is without the use of any caching, which could be used to dramatically reduce this overhead in a production implementation. Figure \ref{fig:resource}, top right and left, show the aggregate NetIO IN and OUT, respectively, over time, indicating that network usage is very stable and predictable.

{\it CPU} indicates the average CPU utilization per container (where 100\% corresponds to full usage of 1 core). Usage over time is also visualized in Figure \ref{fig:resource}, lower right, for 3 nodes, showing that containers oscillate between idling and full usage of 2 cores. CPU usage increases with additional nodes, since nodes need to process more outputs from other nodes and open more connections. However, most of the CPU requirements come from the cryptographic validation operations of the RP, as well as the continuous packet sniffing required to monitor for malicious PPs.

{\it Memory} reports the average amount of RAM each container consumes during operations. Memory requirements increase with additional nodes, as with CPU usage, since nodes need to aggregate more outputs. The memory footprint could be reduced by writing fetched outputs to disk and aggregating one at a time, thus trading off speed for memory. However, the base memory requirements are inherited from the RP, which performs cryptographic validations on all RPKI objects with every validation run.

\begin{table}[h!]
    \centering
    \renewcommand{\arraystretch}{0.9}
    \footnotesize
    \begin{tabular}{r|c|c|c}\hline
         \makecell{\textbf{\# Nodes}} & \makecell{\textbf{NetIO (agg.)} \\ \textbf{IN / OUT}}  & \textbf{CPU (avg.)} & \textbf{Memory (avg.)}  \\ \hline
         3 & 12.4 GB / 9.8 GB & 108\% & 2.5 GB \\
         5 & 17.7 GB / 15.1 GB & 133\% & 2.8 GB \\
         7 & 20 GB / 18 GB & 134\% & 3.2 GB \\
         9 & 24.1 GB / 21.8 GB & 145\% & 3.7 GB \\
         13 & 26.4 GB / 25.4 GB & 140\% & 4.5 GB \\
         15 & 28.2 GB / 27.1 GB & 144\% & 5.0 GB \\\hline
\hline
    \end{tabular}
    \caption{{\small Operation costs per container.}}
    \label{tab:cost}
\end{table}

\subsection{ByzRP: Higher Scalability and Less Traffic}\label{sc:extrapolate}

As more routers use our \brpki\ service we expect the load on the repositories, and consequently failures, to decrease.

We replace $N_{RP} * S_{obj}$ with $N_{RP} * S_{VRP} + N_{node} * S_{obj}$ bytes of traffic, where $N_{RP}$ and $N_{node}$ are the numbers of RPs and nodes, respectively, and $S_{obj}$ and $S_{VRP}$ are the sizes of all RPKI objects and VRPs, respectively. Furthermore, $N_{node} << N_{RP}$ and $S_{VRP} << S_{obj}$. The request volume for PPs would be reduced from $N_{RP}$ to $N_{node}$.

The complete set of RPKI objects across all 64 current PPs constitutes ca. 1.9 GB of data. The VRPs generated from these objects take up about 59 MB in JSON format. However, due to gzip compression by webservers, the inbound network traffic we observe for a single node during one validation run is only about 562 MB, while the compressed VRPs amount to 6.2 MB.

Based on 24 hours of log data collected at our PP, we see RP requests from 3,156 unique subnets. In line with other RPKI measurements\footnote{\url{https://dataplane.org/rpki.html}}, we assume 1 RP per subnet. At the very least this means that $3156 * 562 MB = 1.8 TB$ of traffic in the current RPKI deployment scenario would be reduced to $3156 * 6.2 MB + 15 * 562 MB = 28 GB$ if all RPs were replaced by a 15-node network. That is, \brpki\ would reduce the traffic by a factor of 63. Furthermore, PPs would need to handle only 15 requests per validation round instead of 3156. This means our system could run much shorter validation intervals for faster updates, while still causing less load on the PPs.

The impact of \brpki\ will be even more significant at full RPKI deployment. According to \cite{rovista} 48.2\% of the prefixes are currently covered by ROAs. Extrapolating linearly to 100\% coverage, the (compressed) RPKI data would reach about $562 / 0.482 = 1.2 GB$, while the VRPs would reach $6.2 / 0.482 = 12.9 MB$. In \cite{hlavacek2023beyond} the authors estimate that 70-80K of the roughly 100K ASes on the Internet are non-stub ASes. As the total has increased to 116,702 ASes in April 2024, according to RIR statistics\footnote{\url{https://www-public.imtbs-tsp.eu/~maigron/rir-stats/rir-delegations/world/world-asn-by-number.html}}, we lean towards the higher end at 80K non-stub ASes. In our PP logs we observe that the 3156 subnets belong to 1973 unique ASes, an average of $3156 / 1973 = 1.6$. As a result of these estimates, we calculate the expected traffic volume at full ROA coverage and ROV deployment as $80,000 * 1.6 * 1.2 GB = 153 TB$. A 15-node network at world-wide utilization would reduce this to $80,000 * 1.6 * 12.9 MB + 15 * 1.2 GB = 1.7 TB$.

Both scenarios produce a relative reduction to 1-1.5\% of the original traffic volume. We note here that, due to RRDP and rsync, subsequent updates of the RPKI object cache are much smaller and do not require downloading the entire repository again. However, RTR also supports downloading only new or modified VRPs. Therefore, we expect the relative reduction in traffic to remain similar.

Naturally, the nodes would need to support all of this traffic. However, current PPs are serving full RPKI objects to RPs, thus it would not be unreasonable to assume that the nodes could serve the much smaller VRPs to the same number of RPs, particularly when expanding the network and load-balancing among all nodes. Since routers maintain persistent connections via RTR, this means the RTR servers will need to be equipped with sufficient resources to handle many parallel connections, albeit with very low amounts traffic per connection. So at roughly 80K ASes, an average of 1.6 subnets per AS, and assuming 1 router per subnet, the network would need to support on the order of 128K connections, split across all peers. However, while this requires non-negligible server resources, it is well within modern capabilities. Furthermore, the implementation of a conceptual peer node could be further optimized, e.g., by moving the RTR server into a separate container that can be auto-scaled to meet demand, an wouldn't break the conceptual security guarantees.

\section{Conclusions}\label{sc:conclusions}
Failures in RPKI are expected to increase as RPKI deployment grows. 

Current relying party software packages and deployments are rigid and inflexible, such that errors during validation often lead to full RPKI downgrade \cite{ndss:2024}. Conversely, the protocol itself is also vulnerable to a range of attacks, including those that are difficult to detect because they keep the system deceptively alive, but stop the data fetching and processing by stalling the relying party implementations \cite{van2022rpkiller,stalloris}. Additionally, RPKI publication points have limited resources resulting in connectivity issues to relying parties. The range and effect of vulnerabilities in relying party validators show that they are not stable enough for production environments. 

Failures in RPKI and attacks exploiting vulnerabilities in relying party validators can substantially hinder not only ROV but also the entire slate of protocols for path validation being developed around the RPKI functionality, such as BGPSec and ASPA \cite{aspa,RFC8205}. The effect of attacks and failures will have even more significant consequences as more networks deploy RPKI.

In this work we show how to significantly improve the resilience and security of relying party validation. 

We propose an alternative to individual relying parties: a layer of Byzantine-secure peering relying parties that cooperatively produce a more robust output. The setup can be easily distributed among multiple operators, and since the nodes are designed to achieve consensus with Byzantine fault-tolerance, our proposal does not require trusting individual operators of \brpki\ nodes.
Our infrastructure bypasses the processing and management costs of every AS having to set up, monitor, maintain and upgrade a set of relying parties to supply their BGP routers with updated RPKI data. Instead, our service offers periodically updating VRP sets that can be served directly to routers via the RTR protocol. Furthermore, our infrastructure ensures that even accounting for full RPKI deployment and a significant increase in the size of the RPKI cache, the accompanying network slowdowns, and potential attacks, we can still preserve service availability and generate a comprehensive VRP set within a meaningful time frame, emulating existing RP refresh rates. According to our estimates, \brpki\ can decrease the network load necessary to serve full global RPKI deployment from an expected $153TB$ of data, down to $1.7TB$, a 99\% decrease in load. 

Given that large companies operating inside a specific jurisdiction can be legally coerced to censor networks, an ever expanding \brpki\ network is suitable to facilitate the creation of decentralized volunteer deployments in the same vein as NTP or the TOR network. This system can globally provide up-to-date RPKI data even if some nodes in the network are taken down. Additionally, our proposal can be further improved by deconstructing and rebuilding the relying party processing methods around our infrastructure logic, so that instead of having an external connections monitor, the relying party itself logs and breaks off bad PP connections without the risk of incurring false positive skiplisting.

\section*{Acknowledgements}
This work has been co-funded by the German Federal Ministry of Education and Research and the Hessen State Ministry for Higher Education, Research and Arts within their joint support of the National Research Center for Applied Cybersecurity ATHENE and by the Deutsche Forschungsgemeinschaft (DFG, German Research Foundation) SFB~1119.

\balance

\bibliographystyle{ieeetr}
\bibliography{sec,NetSec,ref,rfc,main,bib}

\end{document}